\pdfoutput=1
\documentclass[aps,prb,preprint,floatfix,groupedaddress,amsmath,amssymb,amsfonts]{revtex4-1}
\usepackage{mathptmx}                    
\usepackage[Symbol]{upgreek}
\usepackage[utf8]{inputenc}
\usepackage{graphicx}
\usepackage[dvipsnames]{xcolor}
\usepackage[
,textwidth=17.5cm
,textheight=23.5cm
,verbose
,pdftex
]{geometry}
\usepackage{xspace}

\usepackage[pdftex]{hyperref}
\hypersetup{
  pdftitle={In/GaN(0001)-(√3×√3)R30° adsorbate structure as a template for embedded (In,Ga)N/GaN monolayers and short-period superlattices},
	colorlinks,
	citecolor=blue
	}

\newcommand{\celsius}{$^{\circ}$C\xspace}
\newcommand{\sqt}{$(\sqrt{3}\times\!\sqrt{3})\text{R}30^{\circ}$\xspace}
\newcommand{\sos}{S$_{1/6}$\xspace}

\DeclareMathAlphabet{\mathsf}{OT1}{\sfdefault}{m}{n}
\SetMathAlphabet{\mathsf}{bold}{OT1}{\sfdefault}{b}{n}

\begin{document}

\title{In/GaN(0001)-$\boldsymbol{{\mathsf{\left(\!\sqrt{3}\times\!\sqrt{3}\right)\!R30^{\circ}}}}$ adsorbate structure as a template for embedded (In,Ga)N/GaN monolayers and short-period superlattices}

\author{C. Chèze}
\email{cheze@pdi-berlin.de}
\author{F. Feix}
\affiliation{Paul-Drude-Institut für Festkörperelektronik, Leibniz-Institut im Forschungsverbund Berlin e.V, Hausvogteiplatz 5--7, 10117 Berlin, Germany}
\author{M. Anikeeva}
\author{T. Schulz}
\author{M. Albrecht}
\affiliation{Leibniz-Institut für Kristallzüchtung, Max-Born-Str.\ 2, 12489 Berlin, Germany}
\author{H. Riechert}
\author{O. Brandt}
\author{R. Calarco}
\affiliation{Paul-Drude-Institut für Festkörperelektronik, Leibniz-Institut im Forschungsverbund Berlin e.V, Hausvogteiplatz 5--7, 10117 Berlin, Germany}

\begin{abstract}
We explore an alternative way to fabricate (In,Ga)N/GaN short-period superlattices on GaN(0001) by plasma-assisted molecular beam epitaxy. We exploit the existence of an In adsorbate structure manifesting itself by a \sqt surface reconstruction observed \textit{in-situ} by reflection high-energy electron diffraction. This In adlayer accommodates a maximum of 1/3 monolayer of In on the GaN surface and, under suitable conditions, can be embedded into GaN to form an In$_{0.33}$Ga$_{0.67}$N quantum sheet whose width is naturally limited to a single monolayer. Periodically inserting these quantum sheets, we synthesize (In,Ga)N/GaN short-period superlattices with abrupt interfaces and high periodicity as demonstrated by x-ray diffractometry and scanning transmission electron microscopy. The embedded quantum sheets are found to consist of single monolayers with an In content of 0.25--0.29. For a barrier thickness of 6 monolayers, the superlattice gives rise to a photoluminescence band at 3.16~eV, close to the theoretically predicted values for these structures.
\end{abstract}


\maketitle
Ternary (In,Ga)N layers are widely used as the active region in optoelectronic devices. The dynamics of charge carriers in these devices is strongly affected by the inevitable compositional fluctuations in the random alloy (In,Ga)N.\cite{Nakamura1998} These fluctuations localize charge carriers, which is beneficial for avoiding nonradiative recombination and thus improves the luminous efficiency of light emitting diodes.\cite{Chichibu2006} However, carrier localization also causes a severe inhomogeneous broadening of electronic transitions and a strongly retarded recombination dynamics, both of which are detrimental for laser operation.\cite{Nakamura1998} In this context, digital (In,Ga)N alloys, i.\,e., heterostructures consisting of an integer number of monolayers (MLs) of the binary compounds InN and GaN are of interest because of the absence of alloy disorder and its consequences.\cite{Lin1994a,Ruterana1999,Cho2005,Yoshikawa2007b}

The fabrication of these digital alloys is challenging due to the large lattice mismatch between InN and GaN (11\%), the tendency of In to segregate on the growth front,\cite{Waltereit2002a} and the incompatible optimum growth conditions of the binary constituents.\cite{Ambacher1998} To achieve visible emission, digital alloys with high average In content are necessary, which in turn requires the fabrication of short-period superlattices (SPSLs) with very thin individual layers. For a well-defined SPSL, abrupt and coherent interfaces between InN and GaN are imperative, and  the incorporation of the accumulated In at the surface must be completely suppressed in GaN. However, for the binary system InN/GaN, plastic relaxation has been found to occur already during the formation of the first InN ML on GaN.\cite{Ng2002,Bellet-Amalric2004}   

A recent detailed investigation of nominal InN/GaN SPSLs by high-resolution transmission electron microscopy (HRTEM) revealed abrupt interfaces, but also showed the structures to actually contain 1~ML thick In$_{0.33}$Ga$_{0.67}$N instead of the expected binary InN layers.\cite{Suski2014} This result suggests that the \sqt-In adsorbate structure on GaN(0001),\cite{Chen2000a, Friedrich2012a} which is established by a third of a ML of In adatoms, plays an important role in the formation of these ultrathin In$_{0.33}$Ga$_{0.67}$N layers.

This In adsorbate structure may in fact constitute an ideal template for the synthesis of laterally ordered InGa$_2$N$_3$ quantum sheets (QSs) with a self-limited thickness of 1~ML. Since its lattice mismatch to GaN only amounts to 3.6\%, plastic relaxation should not be an issue. Furthermore, the In adsorbate structure is stable even at temperatures at which InN decomposes,\cite{Gallinat2007} and hence enables the synthesis of SPSLs at temperatures suitable for the growth of single-crystalline GaN quantum barriers (QBs).

In this Letter, we report the fabrication of (In,Ga)N/GaN SPSLs by taking advantage of the \sqt-In adsorbate structure. Relying on reflection high-energy electron diffraction (RHEED) control, we establish growth conditions promoting the formation of the \sqt-In surface phase and its embedment into a GaN matrix. Employing this fabrication method, we obtain (In,Ga)N/GaN SPSLs with abrupt interfaces. In the present work, we focus on one particular sample (\sos) with ten periods each comprising an (In,Ga)N QS of 1~ML thickness and an In content of 0.25--0.29, close to the maximum one of 0.33, and a GaN QB as thin as 6~MLs. This structure gives rise to a photoluminescence (PL) band at about 3.16~eV coinciding with the theoretically predicted values for SPSLs composed of 1~ML In$_{0.33}$Ga$_{0.67}$N separated by 5--7~MLs GaN.\cite{Suski2014, Gorczyca2015}

\begin{figure*}
\includegraphics*[width=0.85\textwidth]{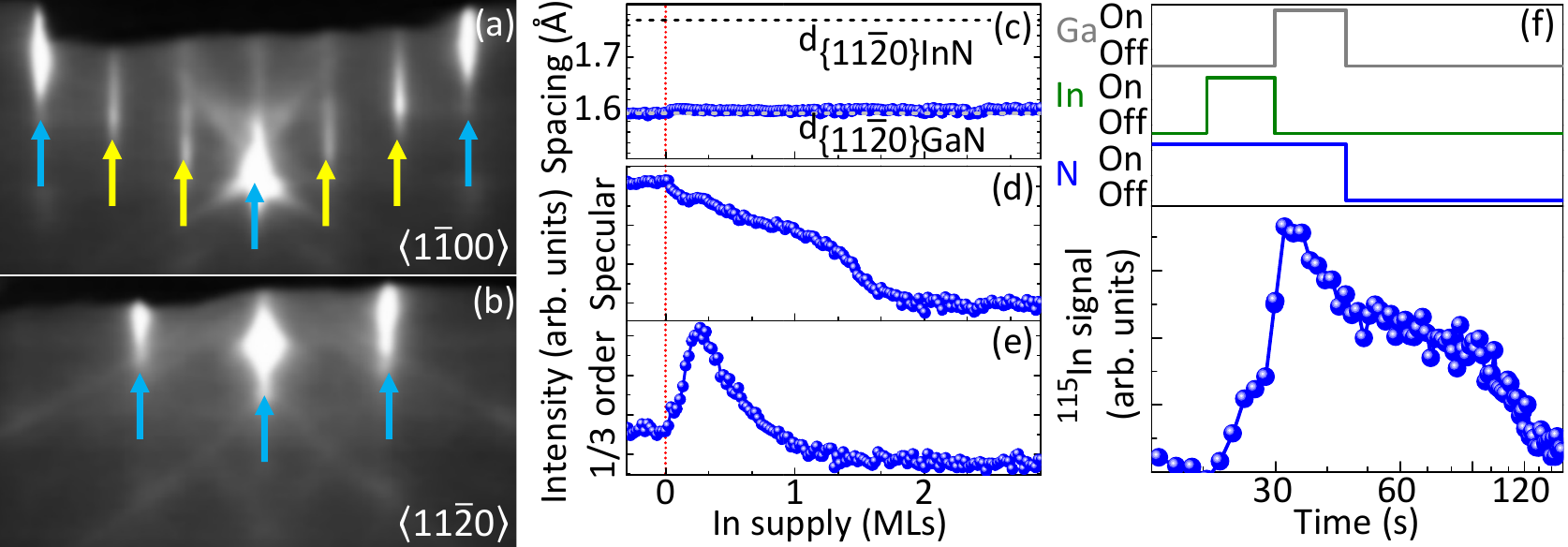}
\caption{(Color online) RHEED patterns along the (a) $\langle1\bar{1}00\rangle$ and (b) $\langle11\bar{2}0\rangle$ azimuths of the GaN(0001) surface showing the In-induced \sqt surface reconstruction for  sample \sos during In and N supply. Blue and yellow arrows indicate integer and fractional order reflections, respectively. Dependence of (c) the $\{11\bar{2}0\}$ spacing (with an uncertainty of $\pm$ 0.2\%), (d) the intensity of the specular reflection, and (e) the intensity of the 1/3 order streak on In supply. (f) $^{115}$In signal monitored by line-of-sight QMS during the formation of the \sqt structure and its overgrowth by GaN in one period of \sos.}
\label{Fig1}
\end{figure*}

The investigation of the \sqt-In structure by RHEED and the synthesis of SPSLs were performed in a DCA Instruments P600  molecular beam epitaxy system equipped with a radio frequency plasma source for generating active N and solid-source effusion cells for Ga and In. Commercial Ga-polar GaN/AlN/Al$_2$O$_3$(0001) templates were used as substrates. The substrate temperature was measured by a pyrometer, which was calibrated using the boundary between the intermediate and the Ga-droplet growth regimes on the GaN(0001) surface.\cite{Heying2000} A substrate temperature of 550~\celsius was established and maintained constant during the entire growth run. The In, Ga, and N fluxes were set to 0.28, 0.37, and 0.34~ML/s, respectively, with 1~ML corresponding to 1.136$\times$10$^{15}$~atoms\,cm$^{-2}$. The surface was exposed to the N flux throughout the deposition process except during a growth interruption following GaN deposition. The In adsorbate structure was formed by flushing the GaN(0001) surface with 2.2~MLs of In, during which an intense \sqt surface reconstruction was observed. This surface was overgrown by 6~MLs of GaN to embed the \sqt-In adsorbate structure in a GaN matrix. Excess In desorbed during GaN growth and the subsequent 455~s growth interruption, as evidenced by line-of-sight quadrupole mass spectrometry (QMS),\cite{Gallinat2007} and was not incorporated into GaN as demonstrated by scanning transmission electron microscopy (STEM) and HRTEM. Hence, the incorporation of In segregating at the growth front into the GaN QBs was suppressed by utilizing Ga-rich conditions.\cite{cheze2016} After the growth interruption, the surface was exposed to N for 20~s to consume excess Ga. This procedure was repeated ten times to form a SPSL.

The RHEED patterns along the perpendicular $\langle1\bar{1}00\rangle$ and $\langle11\bar{2}0\rangle$ azimuths were recorded using a charge-coupled device camera. The intensity of the specular spot and fractional order reflections were measured using a digital RHEED analysis system. The periodicity, interface abruptness and composition of the SPSL under investigation (\sos) were assessed by triple-crystal $\upomega$-$2\uptheta$ x-ray diffraction (XRD) scans performed with CuK$_{\upalpha1}$ radiation in a Panalytical X'Pert\texttrademark\ diffractometer equipped with a Ge(220) hybrid monochromator and a Ge(220) analyzer crystal, as well as by HRTEM and STEM in an aberration-corrected FEI Titan\texttrademark\ 80-300 operating at 300~kV. The distribution of the $c$ lattice parameter was deduced from HRTEM micrographs as described in Ref.~\onlinecite{Schulz2014, Suski2014}. Finally, the spontaneous emission from the SPSLs was evaluated by $\upmu$-photoluminescence spectroscopy (PL) at 10~K using the 325-nm~line of a He-Cd laser for excitation with the laser beam focused to a spot size of about 1~$\upmu$m onto the sample surface.


\begin{figure*}
\includegraphics*[width=0.85\textwidth]{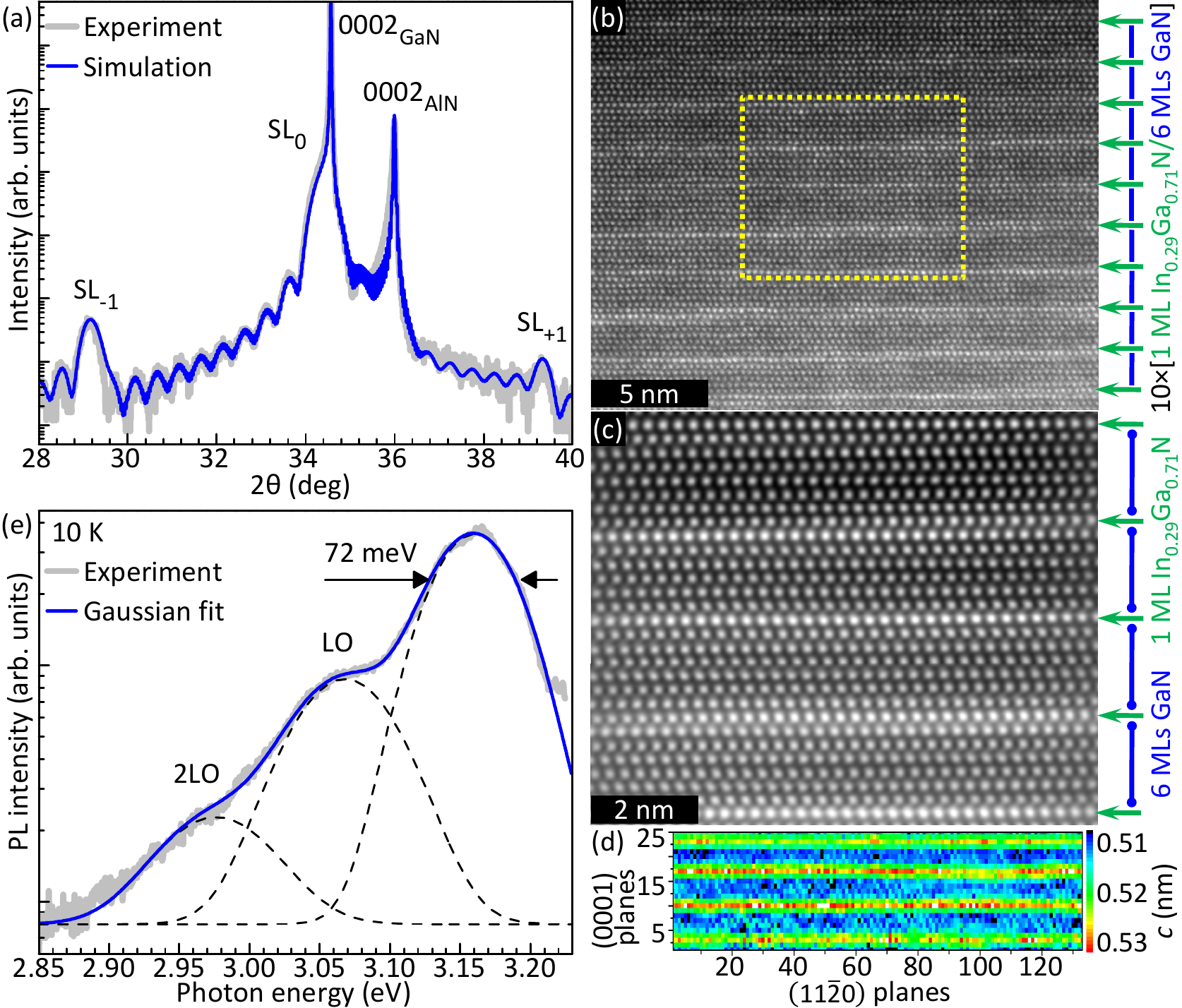}
\caption{(Color online) (a) Experimental and simulated  $\upomega$-$2\uptheta$ XRD profiles of sample \sos across the 0002 reflection of GaN. Satellite reflections due to the SPSL are labelled SL$_{\pm \text{n}}$. (b) Cross-sectional STEM micrograph of the entire SPSL in sample \sos, and (c) magnified view of four periods in the area enclosed by the dashed rectangle in (b). A Fourier filter was carefully applied to the micrograph (c) to avoid artifacts. The horizontal green and vertical blue arrows highlight the (In,Ga)N QSs and the 6~ML thick GaN QBs, respectively. (d) Color-coded map of the $c$ lattice parameter deduced from HRTEM images for four periods of sample \sos revealing a QS thickness of 1~ML and an In content of 25\%. (e) Low-temperature PL spectrum of sample \sos. The blue solid line represents a fit of the spectrum with three Gaussians (dashed lines) separated by 92~meV, the energy of a longitudinal optical (LO) phonon in GaN.}
\label{Fig2}
\end{figure*}

For each period of SPSLs synthesized under the conditions detailed above, the RHEED patterns of the surface taken during the simultaneous supply of In and N exhibit sharp and intense 1/3 order reflections along the $\langle1\bar{1}00\rangle$ azimuth [Fig.~\ref{Fig1}(a)], while only the periodicity of the unreconstructed surface is observed along the $\langle11\bar{2}0\rangle$ azimuth [Fig.~\ref{Fig1}(b)]. These patterns reflect the formation of the \sqt-In adsorbate structure on GaN(0001).

Figures \ref{Fig1}(c)--\ref{Fig1}(e) display the $\{11\bar{2}0\}$ lattice plane distance, the intensity of the specular reflection, and the intensity of the 1/3 order streaks, respectively, during the supply of In and N to the GaN(0001) surface. Figure \ref{Fig1}(c) shows that the in-plane lattice constant equals that of GaN during the entire duration of the In and N supply. This result demonstrates that the high substrate temperature employed inhibits actual InN growth, for which strain relaxation commences within the growth of the first ML, and is almost complete after deposition of 2~MLs.\cite{Ng2002} 

The intensity of the specular reflection [Fig.~\ref{Fig1}(d)] abruptly decreases upon opening the In shutter and stays at a constant value up to a total In supply of 0.32~MLs. Between 0.32 and 1.7~MLs, the intensity monotonically decreases and approaches a steady state at 2.3~MLs. In contrast, the intensity of the 1/3 order streaks shows a clear maximum after a supply of about 0.32~MLs of In [Fig.~\ref{Fig1}(e)]. With continuing In supply, the \sqt reconstruction gets fainter and completely vanishes after about 1.3~MLs. Note that the actual In coverage on the surface is lower than these values due to In desorption.

These results evidence that at a temperature of 550~\celsius and under continuous N supply, In chemisorbs on GaN(0001) and forms ordered and coherent domains of the \sqt-In adsorbate structure. With continued In supply, these domains grow and eventually cover the entire surface with a maximum In coverage of 1/3~ML. In adatoms in excess of this coverage do not give rise to a transition to a different adsorbate structure accommodating a higher In coverage, but accumulate in a liquid metallic phase that merely attenuates the intensity of both the specularly reflected and diffracted beams. Figure \ref{Fig1}(f) shows that during the subsequent deposition of GaN and the following growth interruption, this weakly bonded surplus of In segregates on the growth front and desorbs concurrently as directly monitored by QMS. Note, however, that we have found this excess In to be essential for obtaining embedded (In,Ga)N QSs with an In content close to the maximum expected one, i.\,e., 0.33. Possibly, the liquid In adlayer floating on the surface acts as a reservoir of In that prevents a loss of In from the \sqt-In structure when overgrowth is initiated (for related considerations, see Ref.~\onlinecite{Chen2000b}).  

To examine the structural properties of sample \sos, we employ high-resolution XRD. Figure \ref{Fig2}(a) shows an $\upomega$-$2\uptheta$ scan across the 0002 reflection of GaN. The scan reflects the periodicity of the SPSL since clear first-order satellite reflections are observed. Coherent growth was confirmed by asymmetric $10\bar{1}5$ reciprocal space maps (not shown). With this knowledge, we simulate the XRD profile to obtain the actual thicknesses of the (In,Ga)N and GaN layers as well as the In content in the former. The best fit as shown in Fig.~\ref{Fig2}(a) was obtained with ten periods of 1~ML In$_{0.29}$Ga$_{0.71}$N QSs separated by 5.9~ML GaN QBs.

For imaging the SPSL in real space, we utilize cross-sectional STEM. The micrograph shown in Fig.~\ref{Fig2}(b) was taken from a 50 to 100 nm thick cross-sectional sample and visualizes the periodicity of sample \sos, that is in excellent agreement with the intended one. Most importantly, the interfaces do not deteriorate from the bottom to the top of the structure, but stay abrupt (the progressively weaker contrast toward the top of the SL is due to a change in thickness of the cross-sectional sample). In the magnified view depicted in the micrograph in Fig.~\ref{Fig2}(c), the embedded (In,Ga)N layers seem to exhibit monoatomic steps at some of the (In,Ga)N/GaN interfaces, and the thickness of the (In,Ga)N layers thus appears to fluctuate between 1 and 2~MLs. However, the map of the $c$ lattice parameter deduced from HRTEM on a 10 nm cross-section and shown in Fig.~\ref{Fig2}(d) reveals three increased interplanar distances for each of the (In,Ga)N QS, corresponding to a single ML.\cite{Suski2014} The magnitude of these distortions correspond to an In content of about 25\%, in fair agreement with the one determined from XRD.  

Figure \ref{Fig2}(e) shows a low-temperature PL spectrum of this sample. The emission band from the (In,Ga)N/GaN SPSL peaks at about 3.16~eV and has a full width at half maximum of 72~meV. First and second order longitudinal optical phonon replica are observed at lower energies. The zero-phonon energy of this band is close to the predicted values for SPSLs consisting of 1~ML In$_{0.33}$Ga$_{0.67}$N separated by 5--7~MLs GaN.\cite{Suski2014, Gorczyca2015} Most important, this energy is about 90~meV lower than the one measured for a SPSL also containing 1~ML In$_{0.29}$Ga$_{0.71}$N QSs, but 50~ML GaN QBs.\cite{Feix2016} This redshift reflects the electronic coupling between the individual QSs, which becomes noticeable for barriers thinner than about 7~MLs.\cite{Suski2014, Gorczyca2015} Very similar transition energies have been observed for structures  reportedly consisting of 1~ML InN and 7~MLs GaN,\cite{Kusakabe2015} for which much lower values would be expected theoretically.\cite{Suski2014, Gorczyca2015} Note, however, that the structural parameters of these SPSLs have not been experimentally verified.


The results presented above suggest that the procedure employed here to stabilize and overgrow the \sqt-In adsorbate structure is indeed suitable for embedding it into GaN and thus to reproducibly obtain (In,Ga)N QSs with an In content close to the maximum one of 0.33. Furthermore, the procedure can be repeated periodically for the synthesis of SPSLs, and for a barrier thickness of 6~MLs as realized in the present work, no degradation of interface abruptness is observed even after ten periods. 

The possibility to embed a reconstructed surface phase in a bulk matrix without dissolving it in this matrix may appear as a surprise, since reconstructions are a property of the surface, and are generally not stable in the bulk. However, one has to keep in mind that the \sqt surface reconstruction is primarily a manifestation of an In adsorbate structure on GaN(0001) in the presence of N and should not be confused with, for example, the \sqt surface reconstruction observed for InN(0001) below 350~\celsius.\cite{Himmerlich2009} The exceptional stability of this In adlayer is documented by the fact that we were able to stabilize it up to temperatures of 650~\celsius, about 150~K higher than the dissociation temperature of InN(0001). Note that an intimately related surface phase of the same symmetry is also observed during growth of thick (In,Ga)N(0001) layers,\cite{Chen2000a,Waltereit2002a,Friedrich2012a} which has been attributed to a superstructure of In adatoms in \sqt symmetry on top of an In-depleted subsurface layer.\cite{Friedrich2012a} This self-organized formation of the \sqt-In adsorbate structure demonstrates most clearly that it is the energetically favorable surface phase of In on GaN(0001).

To conclude, we briefly comment on two issues that must be addressed if SPSLs such as fabricated in the present work were to replace conventional (In,Ga)N quantum wells. First, this replacement would only be sensible if the SPSL represents a genuine digital alloy characterized by the absence of alloy disorder. Within the present approach of embedding the \sqt-In structure into GaN, it is thus crucial to ascertain that the high degree of lateral ordering evident from Fig.~\ref{Fig1}(a) can be conserved upon overgrowth.\cite{Gorczyca2015} In a recent study of the carrier recombination dynamics in sample \sos, no evidence for such an ordering was observed. On the contrary, the (In,Ga)N QSs were found to induce a spatially separate localization of electrons and holes analogous to conventional (In,Ga)N QWs.\cite{Feix2016} However, in a two-dimensional system as realized by the (In,Ga)N QSs, any deviation from a perfectly ordered state will induce carrier localization.\cite{Abrahams1979} Detailed microscopic investigations are necessary to elucidate the degree of lateral ordering in the embedded (In,Ga)N QSs, and to devise growth strategies aimed to preserve the initial ordering of the \sqt-In structure. The second important issue concerns the emission wavelength of the SPSL, and thus its average In content. Laser diodes used for Blu-ray\texttrademark\ applications are required to emit at 405~nm (3.06~eV), 90~meV below the transition energy of the sample investigated in the present work, and corresponding to an In content of about 0.1 in the random alloy. To reach this average In content in SPSLs based on the current approach, the width of the GaN QBs has to be decreased to 2 or 3~MLs. Alternatively, while QSs with an In content above 0.33 may be difficult to realize on GaN(0001), it is very well possible that such configurations are stable for other orientations. For example, even a spontaneous ordering into alternating In-rich and Ga-rich MLs has recently been observed for strain-released (In,Ga)N$(000\bar{1})$ fabricated in the form of nanowires.\cite{Zhang2016}


We thank B.~Jenichen for a critical reading of the manuscript and H.\,P.~Schönherr for MBE maintenance. Funding of this work by the European Union's Horizon 2020 research and innovation programme (Marie Skłodowska-Curie Actions) under grant agreement ``SPRInG'' No.\ 642574  and FP7-NMP-2013-SMALL-7 programme under grant agreement No.\ 604416 (DEEPEN) is gratefully acknowledged.

\end{document}